\documentclass{article}
\usepackage{spconf,amsmath,epsfig}
\usepackage{bm}
\usepackage{color}
\usepackage{algorithm}
\usepackage{algorithmic}
\usepackage{amssymb}
\usepackage[justification=centering]{caption}
\usepackage{multirow}
\usepackage{subfigure}
\usepackage{CJK}
\usepackage{epstopdf}
\usepackage{indentfirst}
\usepackage{amsmath}
\usepackage{array,graphicx,subfig}
\usepackage{longtable}
\usepackage{rotating}
\usepackage{multirow}
\usepackage{booktabs}
\usepackage{multirow}
\usepackage{enumerate}
\begin{document}\sloppy

% Example definitions.
% --------------------
\def\x{{\mathbf x}}
\def\L{{\cal L}}

% Title.
% ------
\title{Fusion-supervised Deep Cross-modal Hashing}
%
% Single address.
% ---------------
\name{Li Wang,
        Lei Zhu*,
        En Yu,
        Jiande Sun,
        Huaxiang Zhang*}%
\address{School of Information Science and Engineering, Shandong Normal University\\
Jinan 250358, Shandong Province, China.\\
leizhu0608@gmail.com, huaxzhang@hotmail.com.
}

\maketitle
%\footnote{*Corresponding author}
\let\thefootnote\relax\footnotetext{*Corresponding Author}
\begin{abstract}
Deep hashing has recently received attention in cross-modal retrieval for its impressive advantages. However, existing hashing methods for cross-modal retrieval cannot fully capture the heterogeneous multi-modal correlation and exploit the semantic information. In this paper, we propose a novel \emph{Fusion-supervised Deep Cross-modal Hashing} (FDCH) approach. Firstly, FDCH learns unified binary codes through a fusion hash network with paired samples as input, which effectively enhances the modeling of the correlation of heterogeneous multi-modal data. Then, these high-quality unified hash codes further supervise the training of the modality-specific hash networks for encoding out-of-sample queries. Meanwhile, both pair-wise similarity information and classification information are embedded in the hash networks under one stream framework, which simultaneously preserves cross-modal similarity and keeps semantic consistency. Experimental results on two benchmark datasets demonstrate the state-of-the-art performance of FDCH.
%Furthermore, FDCH keeps the discrete nature of hash codes without relaxation during the optimization.
%However, due to the inability to fully exploit heterogeneous correlation and semantic information, these existing deep cross-modal hashing methods have limited retrieval performance.
\end{abstract}
\begin{keywords}
Cross-modal retrieval, fusion-supervised, heterogeneous modality correlation, pair-wise similarity
\end{keywords}
\begin{figure*}
\centering
{\includegraphics[width=165mm]{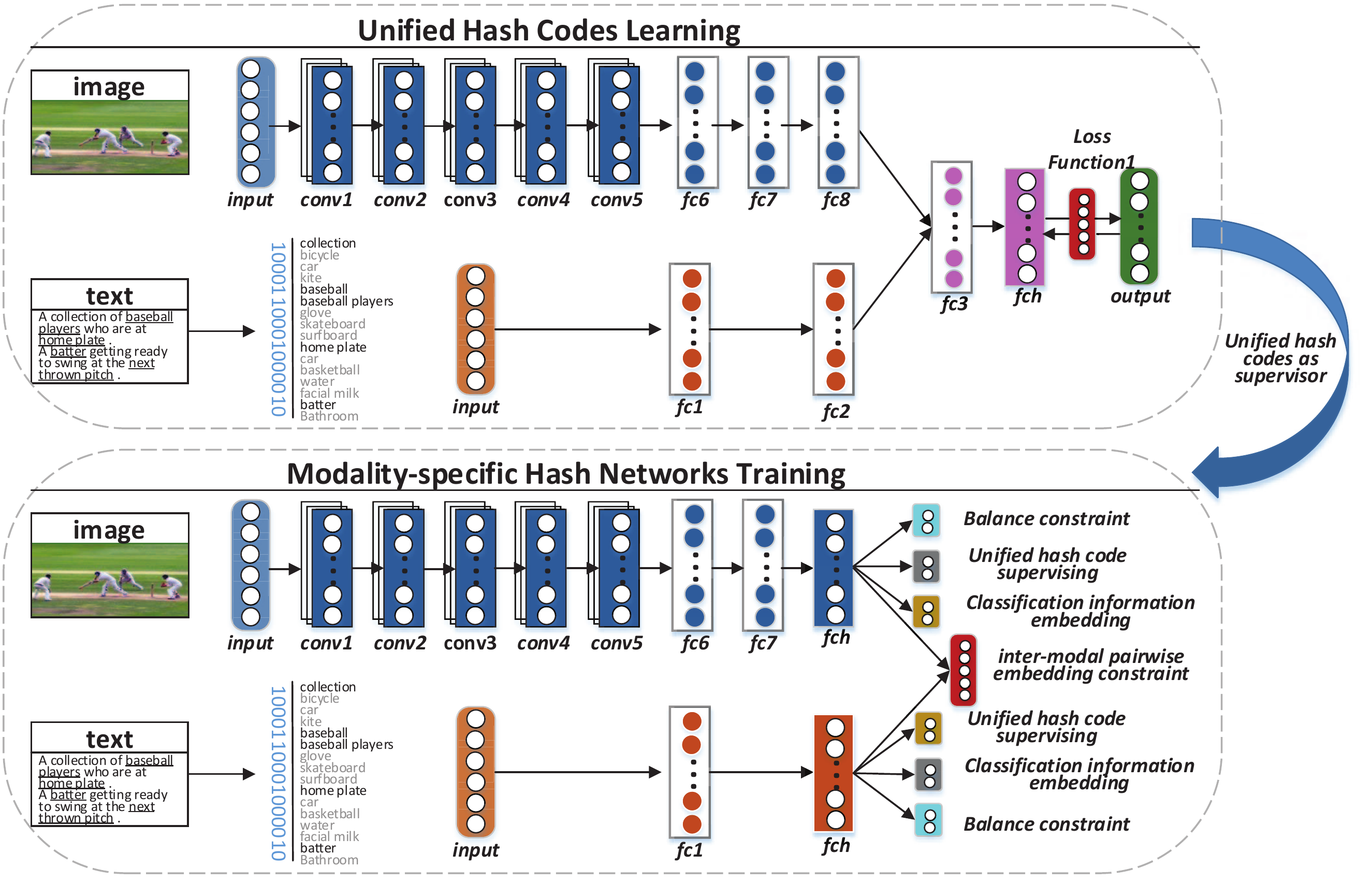}}
%\mbox{\hspace{0.5cm}}
\caption{The deep framework of FDCH model.}
\label{fig:Figure1}
\vspace{-6mm}
\end{figure*}
\section{Introduction}
\label{sec:1}
In recent years, with the rapid growth of different types of data on the Internet, Approximate Nearest Neighbor (ANN) search plays an increasingly important role in many application fields, such as information retrieval, data mining and computer vision \cite{zhu2017exploring,cao2017collective,DBLP:journals/tmm/ZhuHLHSZ17}. Due to the high computational efficiency and low storage cost, hashing has become one of the most popular technologies in ANN search \cite{xie2017dynamic,survey,DBLP:journals/tkde/ZhuSXC17,DBLP:journals/tnn/ZhuHLXS18}.

The basic idea of hashing is to learn the hash function for mapping high-dimensional data into the binary Hamming space while preserving the similarity structure of the original space \cite{survey,zhu2016learning,DBLP:conf/mm/LuZCLNZ19,DBLP:conf/sigir/Lu0CNZ19}. Many hashing-based methods are developed for uni-modal retrieval. Nevertheless, in the real world, data with the same semantic often appears with multiple modalities, such as image, text and video. To take advantage of the relationship between heterogeneous data, it is necessary to develop the Cross-Modal Hashing (CMH) \cite{CMFH,DCH,SCM,SePH,LSSH,xie2016online} in the ANN search. Specifically, in the cross-modal retrieval, the modality of the query is different from that of the retrieved data. In this paper, we mainly focus on two typical cross-modal retrieval tasks: image retrieves text (I2T) and text retrieves image (T2I) tasks.

Most existing CMH methods are based on hand-crafted features. The feature extraction and hash code learning are performed independently. This may limit the representation capability of the samples and thereby undermine the accuracy of the learned hash codes. Recently, the deep learning based hashing methods have developed the end-to-end deep learning architecture to learn feature representation and hash coding simultaneously. Deep Cross-modal Hashing (DCMH) \cite{DCMH} extends the traditional deep models to cross-modal retrieval. It performs the end-to-end learning framework with deep neural networks for each modality. Pairwise Relationship Guided Deep Hashing (PRDH) \cite{PRDH} integrates different pair-wise constraints to learn the similarities of hash codes from inter-modality and intra-modality. With deep learning, deep cross-modal hashing methods can capture the correlation across different modalities more effectively than the shallow learning methods.

In existing deep hashing frameworks, the hash codes from two different modalities of paired samples are compulsively set to be the same. Generally, they first learn the feature representation separately through the deep neural network for different modalities, and then minimize the loss between different modality features to establish the modality relations. The shortcoming of these deep CMH methods is that the complex relationship among multi-modal data cannot be fully captured by simply imposing constraints on the last layer of the neural network with multiple modalities.

In this paper, we propose a novel \emph{Fusion-supervised Deep Cross-modal Hashing} (FDCH) method.
Firstly, the outputs of the deep neural networks for all modalities are imported into the fusion network in pairs. The last layer of the fusion network with pair-wise embedding constraint directly outputs the unified hash codes. By this way, the complex correlation between multi-modal data can be captured more effectively, and the learned unified hash codes enhance the similarity between instances with the same semantic. Furthermore, in order to reduce the quantization error, FDCH directly learns the discrete unified hash codes without relaxing during the optimization. Secondly, we use the unified hash codes as the supervision to train modality-specific hash networks for encoding the out-of-sample data queries. Meanwhile, both pair-wise similarity information and classification information are exploited to train the discriminative hash networks under one stream framework. It enables the learned hash networks to preserve cross-modal similarity and keep semantic consistency simultaneously.

The main contributions of our work are summarized as:
\begin{itemize}
\item FDCH captures the complex correlation between  multi-modal data by inputting the feature representations of paired training samples into a fusion network. Two streams of heterogeneous modalities can be fused deeply. The learned unified hash codes could enhance the similarity between instances with the same semantic, and further supervise the training of the modality-specific hash networks for encoding the queries.
%Besides, these high-quality unified hash codes are used to supervise the training of modality-specific hash networks.\vspace{-2mm}
%FDCH inputs the feature representations of paired training samples into the fusion network to generate the unified hash codes, which further explores the nonlinear correlation of multi-modal data. Besides, these high-quality unified hash codes are used to supervise the training of modality-specific hash networks.
\item We train the modality-specific hash networks under one stream framework with the supervision of both pair-wise similarity information and classification information. Further, FDCH keeps the discrete nature of hash codes without relaxation in the optimization to reduce quantization errors. To the best of our knowledge, there is no similar work in the previous deep cross-modal hashing methods.
%When training the hash networks of image and text, the pair-wise similarity and classification information are embedded under the one stream framework. To the best of our knowledge, there is no similar work in the previous deep cross-modal hashing.\vspace{-2mm}
%\item Unlike most previous CMH methods, FDCH keeps the discrete nature of hash codes without relaxation in the optimization to reduce quantization errors.
%\item Experimental results on two publicly available datasets demonstrate that our FDCH outperforms several state-of-the-art approaches for cross-modal retrieval.
\end{itemize}
%The rest of this paper is organized as follows. First, related work on cross-modal hashing is discussed. Second, the proposed FDCH model and the optimization algorithm are presented. Then the experimental results are shown and analyzed. Finally, we conclude our work in this paper.

\section{Related Work}
\label{sec:2}
Cross-modal hashing has become an active research topic in literatures due to its high efficiency and low storage cost in cross-modal retrieval \cite{CMDVH,Shen2015Learning,DBLP:conf/mm/LiuNZCZY18,DBLP:journals/sigpro/LuZCSZ19}. Various techniques have been proposed for CMH. They can be roughly divided into two categories: unsupervised methods and supervised methods.%large-scale

Unsupervised hashing methods learn the hash codes and functions with only the paired unlabeled training samples. Inter-Media Hashing (IMH) \cite{IMH} finds a common Hamming space for learning hash functions, where inter-modality and intra-modality consistency are explored to produce hash codes. Linear Cross-Modal Hashing (LCMH) \cite{LCMH} learns the common space and hash functions by representing training samples from all modalities with lower dimensional approximations. It preserves the similarity of multimedia documents and reduces the time and space complexity. Collective Matrix Factorization Hashing (CMFH) \cite{CMFH} learns hash codes with collective matrix factorization during the offline phase. Latent Semantic Sparse Hashing (LSSH) \cite{LSSH} learns the latent semantic features through sparse coding and matrix factorization, and then maps them to a joint abstraction space to obtain the binary hash codes.

Supervised hashing methods achieve better results by exploiting semantic information to enhance correlation across different modalities.
Semantic Correlation Maximization (SCM) \cite{SCM} maximizes the correlation between different modalities by seamlessly integrating semantic label into the hashing learning for large-scale data. Semantics-Preserving Hashing (SePH) \cite{SePH} learns the semantics-preserving hash codes by minimizing the KL-divergence of derived probability distribution in the hamming space from that in semantic space. Discriminant Cross-modal Hashing (DCH) \cite{DCH} is proposed to jointly learn the binary codes, the linear classifiers and the hash functions.

The methods mentioned above are all shallow CMH methods based on hand-crafted features. Since the feature extraction and hash code learning processes are performed separately, they may not interact to further exploit the relationships among different modalities.
Recently, several works on deep CMH have been developed to perform cross-modal retrieval through the end-to-end learning architecture with deep neural networks \cite{DCMH,PRDH}. %Representative methods include Deep Cross-modal Hashing (DCMH) \cite{DCMH}, Deep Visual-Semantic Hashing (DVSH) \cite{DVSH}, Pairwise Relationship Guided Deep Hashing (PRDH) \cite{PRDH} and Dual Deep Neural Networks Cross-Modal Hashing (DDCMH) \cite{DDCMH}.
Nevertheless, these methods are not able to fully capture the nonlinear correlation across different modalities to learn discriminative hash codes.

\section{The proposed approach}
\label{sec:3}
%In this section, we first introduce the deep framework overview and problem formulation. Next, since FSDH is a two-step deep hashing method, its training process is divided into two steps, including unified hash codes learning and modality-specific hash networks training. Finally, the details of out-of-sample extension are described.
\subsection{Deep Framework Overview}
\label{sec:3.1}
 The basic deep framework of FDCH model is illustrated in  Figure \ref{fig:Figure1}. It consists of two components: %unified hash codes learning and modality-specific hash networks training.

\textbf{\emph{Unified hash codes learning}}: There are three networks involved in this component: image network, text network and fusion network. For the design of the image network, we use the CNN-F network in \cite{Chatfield2014Return}. The original CNN-F model consists of eight layers: five convolutional layers $(conv1-conv5)$ and three fully-connected layers $(fc6-fc8)$.
For the text modality, we first represent each text sample as a vector with Bag-of-Word (BOW) representation, and then input the BOW vector into the text network with two fully-connected layers. In particular, the number of hidden nodes in the last fully-connected layer of the image and text network should be consistent.
For fusion network combines the outputs of image and text networks in pairs, it consists of two fully-connected layers. To obtain the unified hash codes, the second layer is designed as a hash layer $fch$ (with $k$ hidden nodes), and the activation function is an identity function.

\textbf{\emph{Modality-specific hash networks training}}: In order to learn the hash functions for encoding out-of-sample queries, we redesign the image and text networks for training modality-specific hash networks. In this component, the last fully-connected layers of the image and text networks are replaced with the hash layer $fch$ (with $k$ hidden nodes), the identity function is used as the activation function, the rest of the settings are the same as the above step.%\vspace{-2mm}

\subsection{Notations and Problem Formulation}
\label{sec:3.2}
Assume cross-modal dataset is $O = \left\{ {\left. {{o_i}} \right\}} \right._{i = 1}^n$ containing $n$ training sample pairs, where ${o_i} = \left( {{v_i},{t_i},{y_i}} \right)$. Correspondingly, $V = \left[ {{v_1},...,{v_n}} \right] \in {\mathbb{R}^{{d_v} \times n}}$ and $T = \left[ {{t_1},...,{t_n}} \right] \in {\mathbb{R}^{{d_t} \times n}}$ are image and text features, respectively. Let $Y = \left [{{y_1},...,{y_n}} \right] \in {\mathbb{R}^{{c} \times n}}$ denote label matrix for all training samples, and if $o_i $ belongs to the $j$-th class ${y_{ji}} = 1$, otherwise ${y_{ji}} = 0$. Given the label information, the pair-wise similarity matrix is defined as $S = \left\{ {\left. {{s_{ij}}} \right\}} \right.$, where ${s_{ij}} = 1$ when ${o_i}$ is semantically similar to ${o_j}$, otherwise ${s_{ij}}=0$. For multi-label datasets, we define ${s_{ij}} = 1$ if ${o_i}$ and ${o_j}$ have at least one same label, otherwise ${s_{ij}}=0$.

FDCH aims to learn compact $k$-bit binary code for each instance, and $B \in {\left\{ { - 1,1} \right\}^{k \times n}}$ is unified hash code matrix with the $i$-th column ${b_i} \in {\left\{ { - 1,1} \right\}^k}$. For two binary codes ${b_i}$ and ${b_j}$, the relationship between their Hamming distances $di{s_H}\left( {{b_i},{b_j}} \right)$ for evaluating similarity and their inner product $\left\langle {{b_i},{b_j}} \right\rangle $ is formulated as follows: $di{s_H}\left( {{b_i},{b_j}} \right) = \frac{1}{2}\left( {k - \left\langle {{b_i},{b_j}} \right\rangle } \right)$. Obviously, two binary codes with larger inner products have smaller Hamming distances. Therefore, we can use the inner product of different binary codes to quantify their similarity.
Given pair-wise similarity matrix $S = \left\{ {\left. {{s_{ij}}} \right\}} \right.$, the probability of ${s_{ij}}$ under the condition $B$ is defined as:
\begin{equation}
\small
\label{eq:1}
\begin{split}
p\left( {{s_{ij}}|B} \right) = \left\{ \begin{array}{l}
\sigma \left( {{\Psi _{ij}}} \right),{s_{ij}} = 1\\
1 - \sigma \left( {{\Psi _{ij}}} \right),{s_{ij}} = 0
\end{array} \right.
\end{split}
\end{equation}
where $\sigma \left( {{\Psi _{ij}}} \right) = \frac{1}{{1 + {e^{ - {\Psi _{ij}}}}}}$ is a sigmoid function, and ${\Psi _{ij}} = \frac{1}{2}\left\langle {{b_i},{b_j}} \right\rangle  = \frac{1}{2}b_i^T{b_j}$. Eq.(\ref{eq:1}) shows that samples with larger inner products should have higher probability of being similar, so it can measure the similarity of hash codes.%\vspace{-2mm}

\subsection{Unified Hash Codes Learning}
\label{sec:3.3}
Let $\mathop f\limits^ \sim  \left( {V;{\theta _v}} \right)$ denote the learned feature representation corresponding to the output of the image network, $\mathop g\limits^ \sim  \left( {T;{\theta _t}} \right)$ denote the learned text feature corresponding to the output of the text network. Given the output of the image and text networks, we can combine them with the nonlinear activation (tanh) function $t\left(  \cdot  \right)$ to get the input $Z = t\left( {} \right.\mathop f\limits^ \sim  \left( {V;{\theta _v}} \right) + \mathop g\limits^ \sim  \left( {T;{\theta _t}} \right)\left. {} \right)$ of the fusion network. Further, we define the output of the fusion network for all training as $H = h\left( {Z;{\theta _z}} \right) \in {\mathbb{R}^{K \times n}}$.
By jointly training the image network, text network and fusion network, our method nonlinearly transforms the paired training samples into powerful unified hash codes in this step.

Combining with the above analysis of similarity measure, the objective function for learning the unified hash codes can be defined as follows
%To learn unified hash codes, the objective function can be defined as follows:
\begin{equation}
\small
\label{eq:2}
\begin{split}
\begin{array}{l}
\mathop {\min }\limits_{B,{\theta _v},{\theta _t},{\theta _z}} L =  - \sum\limits_{i,j = 1}^n {\log p\left( {{s_{ij}}|B} \right)} \\
\qquad \qquad \qquad + \lambda \left\| {B - H} \right\|_F^2 + \eta \left\| {H\textbf{1}} \right\|_F^2\\
\qquad \qquad \qquad s.t.B \in {\left\{ { - 1,1} \right\}^{k \times n}},
\end{array}
\end{split}
\end{equation}
where the first term is a negative log likelihood function $-\sum\limits_{{s_{ij}} \in S} {\log p\left( {{s_{ij}}|B} \right)}  =  - \sum\limits_{{s_{ij}} \in S} {\left( {{s_{ij}}{\Phi _{ij}} - \log \left( {1 + {e^{{\Phi _{ij}}}}} \right)} \right)} $, and ${\Phi _{ij}} = \frac{1}{2}H_{ * i}^T{H_{ * j}}$. Besides, $\lambda $ and $\eta$ are hyper-parameters.
The first term is a pair-wise embedding constraint to preserve the semantic similarities in $S$. By minimizing the negative log likelihood function, it makes the similarity (inner product) between two similar samples as large as possible, and the similarity between dissimilar samples as small as possible. The second term minimizes the loss between outputs of the fusion network and binary hash codes, so that the learned unified hash codes can preserve the nonlinear correlations among training samples. The third term is a balance constraint to maximize the information of each hash bit. It requires each bit to have an equal opportunity to be coded to 1 or -1 for all training samples.
 %The third constraint is imposed to maximize the information from each hash bit, which occurs when each bit leads to a balanced partitioning of the dataset.

The optimization problem in Eq.(\ref{eq:2}) can be solved by using the iterative update strategy. Specifically, we can learn one variable while fix other variables alternatively. We learn unified hash codes $B$ and network parameters $\theta=\left\{ {{\theta _v},{\theta _t},{\theta _z}} \right\}$ as follows.

\textbf{\emph{Update $\theta $, with $B$ fixed}}: we adopt the mini-batch Stochastic Gradient Descent (SGD) to learn $\theta  = \left\{ {{\theta _v},{\theta _t},{\theta _z}} \right\}$. We first compute the gradient of the loss function in Eq.(\ref{eq:2})
\begin{equation}
\small
\label{eq:3}
\begin{split}
\begin{array}{l}
\frac{{\partial L}}{{\partial {H_{ * i}}}} = \frac{1}{2}\sum\limits_{j = 1}^n {\left( {\sigma \left( {{\Phi _{ij}}} \right){H_{ * j}} - {s_{ij}}{H_{ * j}}} \right)} \\
\qquad \quad + \frac{1}{2}\sum\limits_{j = 1}^n {\left( {\sigma \left( {{\Phi _{ji}}} \right){H_{ * j}} - {s_{ji}}{H_{ * j}}} \right)} \\
\qquad \quad + 2\lambda \left( {{H_{ * i}} - {B_{ * i}}} \right) + 2\eta H\textbf{1}
\end{array}
\end{split}
\end{equation}

Then we can obtain the gradients of each layer through the chain rule and update the parameters $\theta$ by Back Propagation (BP) \cite{hinton2006reducing} algorithm.

\textbf{\emph{Update $B$, with $\theta $ fixed}}: In the optimization process, we keep the discrete nature of unified hash codes $B$. When $\theta $ is fixed, we can obtain the following formulation
\begin{equation}
\small
\label{eq:4}
\begin{split}
\begin{array}{l}
\mathop {\max }\limits_B tr\left( {\lambda {B^T}H} \right) \ s.t. \ B \in {\left\{ { - 1,1} \right\}^{k \times n}}
\end{array}
\end{split}
\end{equation}

The hash codes ${B_{ij}}$ can be solved as
\begin{equation}
\small
\label{eq:5}
\begin{split}
B = \texttt{sign}\left( {\lambda H} \right)
\end{split}
\end{equation}
\subsection{Modality-specific Hash Networks Training}
\label{sec:3.4}
We have learned powerful unified hash codes for all training sample pairs through above steps. In this section, we train the image network $f\left( {V;{\theta _v}} \right)$ and text network $g\left( {T;{\theta _t}} \right)$ to obtain corresponding hash functions ${h^v}\left(  \cdot  \right)$ and ${h^t}\left(  \cdot  \right)$ for encoding out-of-sample queries.

Firstly, we adopt the inter-modal pair-wise embedding constraint preserve the cross-modal similarity between the outputs from image network hash and text hash network.
\begin{equation}
\small
\label{eq:6}
\begin{split}
{J_1} =  - \sum\limits_{i,j = 1}^n {\left( {{s_{ij}}{\Theta _{ij}} - \log \left( {1 + {e^{{\Theta _{ij}}}}} \right)} \right)}
\end{split}
\end{equation}
where ${\Theta _{ij}} = F_{ * i}^T{G_{ * j}}$.

Obviously, by optimizing the negative log likelihood of the pair-wise similarity $S$, the Hamming distance between two semantically similar samples can be reduced, while increasing the Hamming distance between semantically dissimilar samples.

Then, we use the high-quality hash codes learned before to supervise the training of modality-specific hash networks. By minimizing the loss function below, the nonlinear correlations embedded in $B$ are transmitted to hash networks $f\left( {V;{\theta _v}} \right)$ and $g\left( {T;{\theta _t}} \right)$. At the same time, the outputs of the last fully connected layer for two hash networks are forced to be close to the binary code.
\begin{equation}
\small
\label{eq:7}
\begin{split}
{J_2} = \left\| {B - F} \right\|_F^2 + \left\| {B - G} \right\|_F^2
\end{split}
\end{equation}

Although the pair-wise similarity information in Eq.(\ref{eq:6}) and the unified hash codes containing the nonlinear correlation in Eq.(\ref{eq:7}) have been used to learn the hash functions, the semantic label information is not fully exploited.

Next, we model the relationship between the hash networks and the label information. The trained hash networks of image and text are expected to be optimal for classification. We use the label information directly by linearly mapping it into modality-specific networks as follows
\begin{equation}
\small
\label{eq:8}
\begin{split}
\begin{array}{l}
{J_3} = \left\| {F - W_1^TY} \right\|_F^2 + \left\| {G - W_2^TY} \right\|_F^2 + \left\| {W_1^T} \right\|_F^2 + \left\| {W_2^\texttt{T}} \right\|_F^2
\end{array}
\end{split}
\end{equation}
where $W_1^\texttt{T} \in {\mathbb{R}^{k \times c}}$ and $W_2^\texttt{T} \in {\mathbb{R}^{k \times c}}$ are mapping matrices corresponding to image and text modalities, respectively.

Finally, we add the following balance constraints to maximize the information of each bit.
\begin{equation}
\small
\label{eq:9}
\begin{split}
\begin{array}{l}
{J_4} = \left\| {F\textbf{1}} \right\|_F^2 + \left\| {G\textbf{1}} \right\|_F^2
\end{array}
\end{split}
\end{equation}

Based on the analysis of the above four terms, the overall objective function is rewritten as follows
\begin{equation}
\small
\label{eq:10}
\begin{split}
\mathop {\min }\limits_{{\theta _v},{\theta _t}} J = {J_1} + \gamma {J_2} + \beta {J_3} + \alpha {J_4}
\end{split}
\end{equation}
where $\gamma $, $\beta $ and $\alpha $ are hyper-parameters to balance the importance of each terms.
\begin{table*}[]
\small
\caption{mAP comparison on MIRFLICKR-25K and NUS-WIDE. The best result is shown in boldface.}
\vspace{-3mm}
\centering
\resizebox{1.5\columnwidth}{!}{
\begin{tabular}{|p{25mm}<{\centering}|p{14mm}<{}|p{12mm}<{\centering}|p{12mm}<{\centering}|p{12mm}<{\centering}|p{12mm}<{\centering}|p{12mm}<{\centering}|p{12mm}<{\centering}|}
%\begin{tabular}{|c|l|l|l|l|l|l|l|}
\hline
\multirow{2}{*}{Task}  &\multicolumn{1}{c|}{\multirow{2}{*}{Methods}}  &\multicolumn{3}{c|}{\textbf{MIRFLICKR-25K}}   &\multicolumn{3}{c|}{\textbf{NUS-WIDE}} \\ \cline{3-8}
& \multicolumn{1}{c|}{}                         & 16 bit     & 32 bit     & 64 bit     & 16 bit     & 32 bit     & 64 bit     \\ \hline \hline
\multirow{3}{*}{\begin{tabular}[c]{@{}c@{}} \\ \\  Image Query\\ v.s\\ Text Database\end{tabular}}
 & CMFH               & 0.5644       & 0.5648          & 0.5666          & 0.3384       & 0.3407          & 0.3452         \\ \cline{2-8}
 & LSSH               & 0.5950       & 0.6043          & 0.6117          & 0.4616       & 0.4606          & 0.4800          \\ \cline{2-8}
 & DCH                & 0.5991       & 0.6025          & 0.6009          & 0.4275       & 0.4402          & 0.4526          \\ \cline{2-8}
 & SCM                & 0.5821       & 0.5726          & 0.5672          & 0.3858       & 0.3744          & 0.3666          \\ \cline{2-8}
 & SePH$_{km}$        & 0.6534       & 0.6665          & 0.7277          & 0.6056       & 0.6223          & 0.6674          \\ \cline{2-8}
 & DCMH               & 0.7050       & 0.7194          & 0.7366          & 0.6195       & 0.6472          & 0.6721          \\ \cline{2-8}
 %& PRDH               & 0.7126       & 0.7128          & 0.7201          & 0.6348       & 0.6529          & 0.6506          \\ \cline{2-8}
 & FDCH              & \textbf{0.7762}       & \textbf{0.7877}         & \textbf{0.7897}     & \textbf{0.6861}       & \textbf{0.7003}         & \textbf{0.7139 }  \\ \hline \hline
\multirow{3}{*}{\begin{tabular}[c]{@{}c@{}}\\ \\ Text Query\\ v.s\\ Image Database\end{tabular}}
 & CMFH               & 0.5632       & 0.5631          & 0.5639          & 0.3377       & 0.3401          & 0.3424          \\ \cline{2-8}
 & LSSH               & 0.5893       & 0.5988          & 0.5950          & 0.4481       & 0.4464          & 0.4568          \\ \cline{2-8}
 & DCH                & 0.5804       & 0.5981          & 0.5968          & 0.4003       & 0.4179          & 0.4428          \\ \cline{2-8}
 & SCM                & 0.5976       & 0.5826          & 0.5724          & 0.3858       & 0.3708          & 0.3576          \\ \cline{2-8}
 & SePH$_{km}$        & 0.6135       & 0.6238          & 0.6768          & 0.5876       & 0.5747          & 0.6314          \\ \cline{2-8}
 & DCMH               & 0.7038       & 0.7048          & 0.7291          & 0.6147       & 0.6432          & 0.6651          \\ \cline{2-8}
 %& PRDH               & 0.7467       & 0.7540          & 0.7505          & 0.6808       & 0.6961          & 0.6943         \\ \cline{2-8}
 & FDCH              & \textbf{0.7504 }      &\textbf{0.7555 }       & \textbf{0.7552}     & \textbf{0.6518 }      & \textbf{0.6712}         & \textbf{0.6834 }     \\ \hline
\end{tabular}
}
\label{table2}
\vspace{-3mm}
\end{table*}

Similarly, for the optimization problem of Eq.(\ref{eq:10}), we can also solve it through the alternate learning strategy, which learns one parameter by fixing others. The process of alternately learning parameters $\theta _v$, $\theta _t$, $W_1$ and $W_2$ is as follows.

\textbf{\emph{Update $\theta_v $, with $\theta _t$, $W_1$ and $W_2$ fixed}}: The mini-batch SGD with BP algorithm is used to learn parameter $\theta _v$. For each instance ${{v_i}}$, we first derive the gradient of the loss function as follow
\begin{equation}
\small
\label{eq:11}
\begin{split}
\begin{array}{l}
\frac{{\partial J}}{{\partial {F_{ * i}}}} = \frac{1}{2}\sum\limits_{j = 1}^n {\left( {\sigma \left( {{\Theta _{ij}}} \right){G_{ * j}} - {s_{ij}}{G_{ * j}}} \right)}+ 2\gamma \left( {{F_{ * i}} - {B_{ * i}}} \right) \\
\qquad \quad+ 2\beta \left( {W_1^\texttt{T}{Y_{ * i}} - {F_{ * i}}} \right) + 2\alpha F\textbf{1}
\end{array}
\end{split}
\end{equation}

Then we can obtain the gradients of each layer through the chain rule and update the parameters $\theta_v $ by BP algorithm.

\textbf{\emph{Update $\theta_t $, with $\theta _v$, $W_1$ and $W_2$ fixed}}: Similarly, we learn parameter $\theta _t$ through the mini-batch SGD with BP algorithm. For each instance ${{v_t}}$, we compute the gradient of the loss function as follow
\begin{equation}
\small
\label{eq:12}
\begin{split}
\begin{array}{l}
\frac{{\partial J}}{{\partial {G_{ * j}}}} = \frac{1}{2}\sum\limits_{i = 1}^n {\left( {\sigma \left( {{\Theta _{ij}}} \right){F_{ * i}} - {s_{ij}}{F_{ * i}}} \right)}+ 2\gamma \left( {{G_{ * j}} - {B_{ * i}}} \right) \\
\qquad \quad + 2\beta \left( {W_2^\texttt{T}{Y_{ * j}} - {G_{ * j}}} \right) + 2\alpha G\textbf{1}
\end{array}
\end{split}
\end{equation}

Then we can obtain the gradients of each layer through the chain rule and update the parameters $\theta_t $ by BP algorithm.

\textbf{\emph{Update $W_1$, with $\theta _v$, $\theta_t $ and $W_2$ fixed}}: It is easy to solve parameter $W_1$ by the regularized least squares problem. We can get the following closed-form solution
\begin{equation}
\small
\label{eq:13}
\begin{split}
W_1 = \beta {\left( {Y{Y^\texttt{T}} + I} \right)^{ - 1}}Y{F^\texttt{T}}
\end{split}
\end{equation}

\textbf{\emph{Update $W_2$, with $\theta _v$, $\theta_t $ and $W_1$ fixed}}: The parameter $W_2$ can also be solved by the regularized least squares problem. It has a closed-form solution as follow:
\begin{equation}
\small
\label{eq:14}
\begin{split}
W_2 = \beta {\left( {Y{Y^\texttt{T}} + I} \right)^{ - 1}}Y{G^\texttt{T}}
\end{split}
\end{equation}

\subsection{Online Cross-modal Retrieval}
For a new query that is out of the retrieval database, we can easily obtain its hash code as long as one of its modalities is available. Specifically, given a query instance ${v_q}$ from the image modality, we use forward propagation of the image network to generate the hash code as follows:%use it as an input of the image network
\begin{equation}
\small
\label{eq:15}
\begin{split}
b_q^v = {h^v}\left( {{v_q}} \right) = \texttt{sign}\left( {f\left( {{v_q};{\theta _v}} \right)} \right)
\end{split}
\end{equation}

Similarly, if a instance only has the text modality ${t_q}$, we can also use the text network to generate its hash code as:
\begin{equation}
\small
\label{eq:16}
\begin{split}
b_q^t = {h^t}\left( {{t_q}} \right) = \texttt{sign}\left( {g\left( {{t_q};{\theta _t}} \right)} \right)
\end{split}
\end{equation}

\section{Experiments}
\label{sec:4}
%To evaluate the performance of FDCH, we conducted extensive experiments on two public image-text datasets and compared FDCH with several state-of-the-art cross-modal hashing methods. The following sections describe the details of the experiment. \vspace{-10mm}

\subsection{Experimental Datasets}
\label{sec:4.1}
\textbf{MIRFLICKR-25K} \cite{Huiskes2008The} consists of 25,000 instances collected from Flickr website. In experiments, we select instances with at least 20 labels, which has 20,015 image-text pairs. Here, each textual modality is represented as a 1386-dimensional BoW vector. For image modality, we directly use raw pixels as input. %In experiments, we randomly sample 2,000 instances as query and the rest as database. Similar to \cite{PRDH}, we randomly sample 5,000 instances from the database for training.
%, each of which contains an image and several associated textual tags.

\textbf{NUS-WIDE} \cite{Chua2009NUS} is a real-world web image dataset containing 269,648 instances. In the experiment, we select the 10 most frequent concepts to construct a subset with 186,577 image-text pairs. For each image-text pair, the textual modality is represented by 1000-dimensional BoW vector, and the image modality directly uses the raw pixels as input. %On this dataset, we randomly sample about 1\% instances as query and the rest as database. Similarly, 5000 data samples are randomly sampled from the database for training.

In experiments, we randomly sample 2,000 instances as query and the rest as database (retrieval set). To reduce computational cost, similar to \cite{PRDH}, we randomly sample 5,000 instances from the database for training. After completing the training, we binarize the database samples into the hash codes and perform cross-modal retrieval. %Therefore, our approach is scalable for large datasets.
\vspace{-4mm}
\begin{figure*}
\centering
\mbox{
\subfigure{\includegraphics[width=70mm]{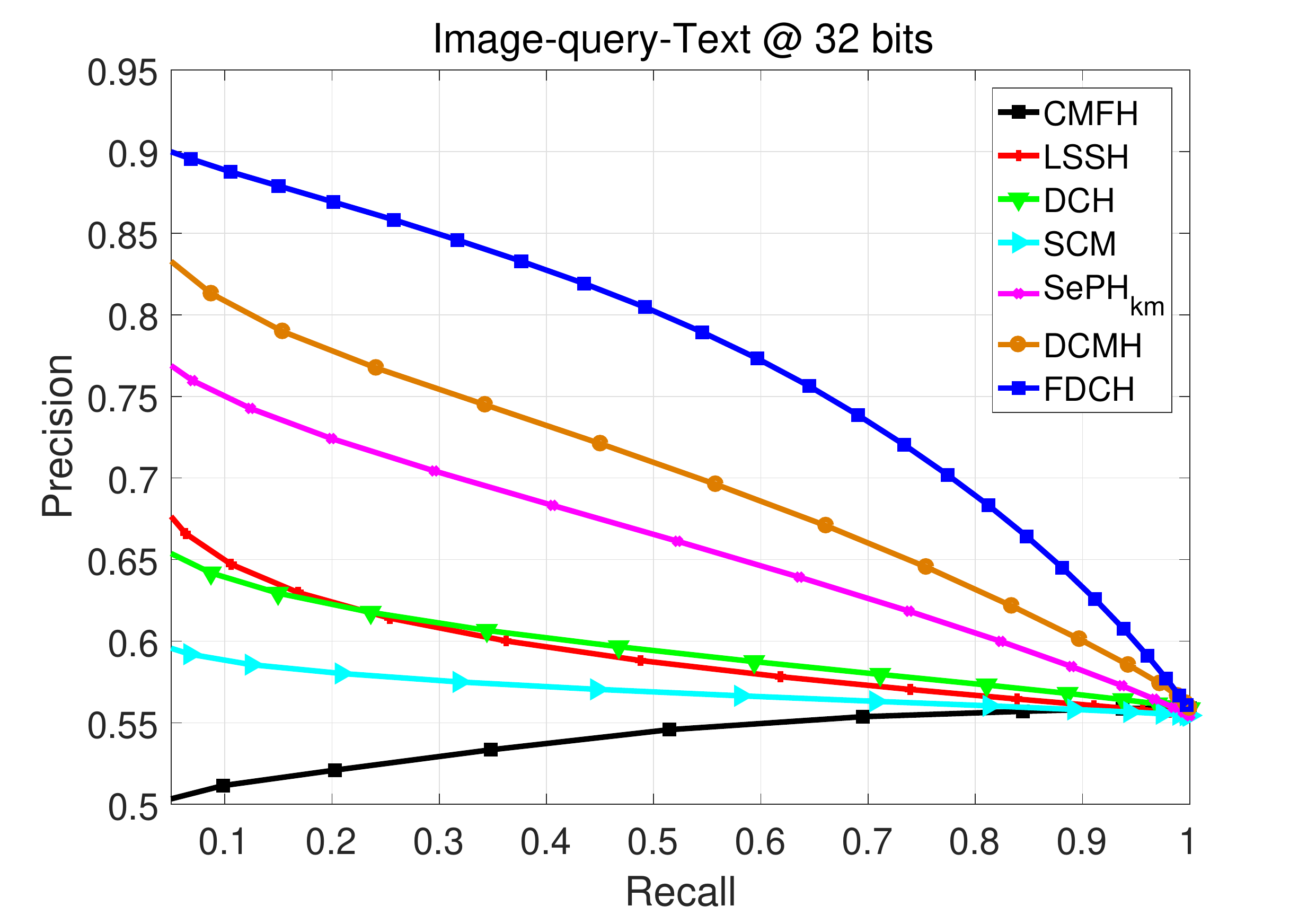}}% \hspace{-5mm}
}\hspace{-4mm}\mbox{
\subfigure{\includegraphics[width=70mm]{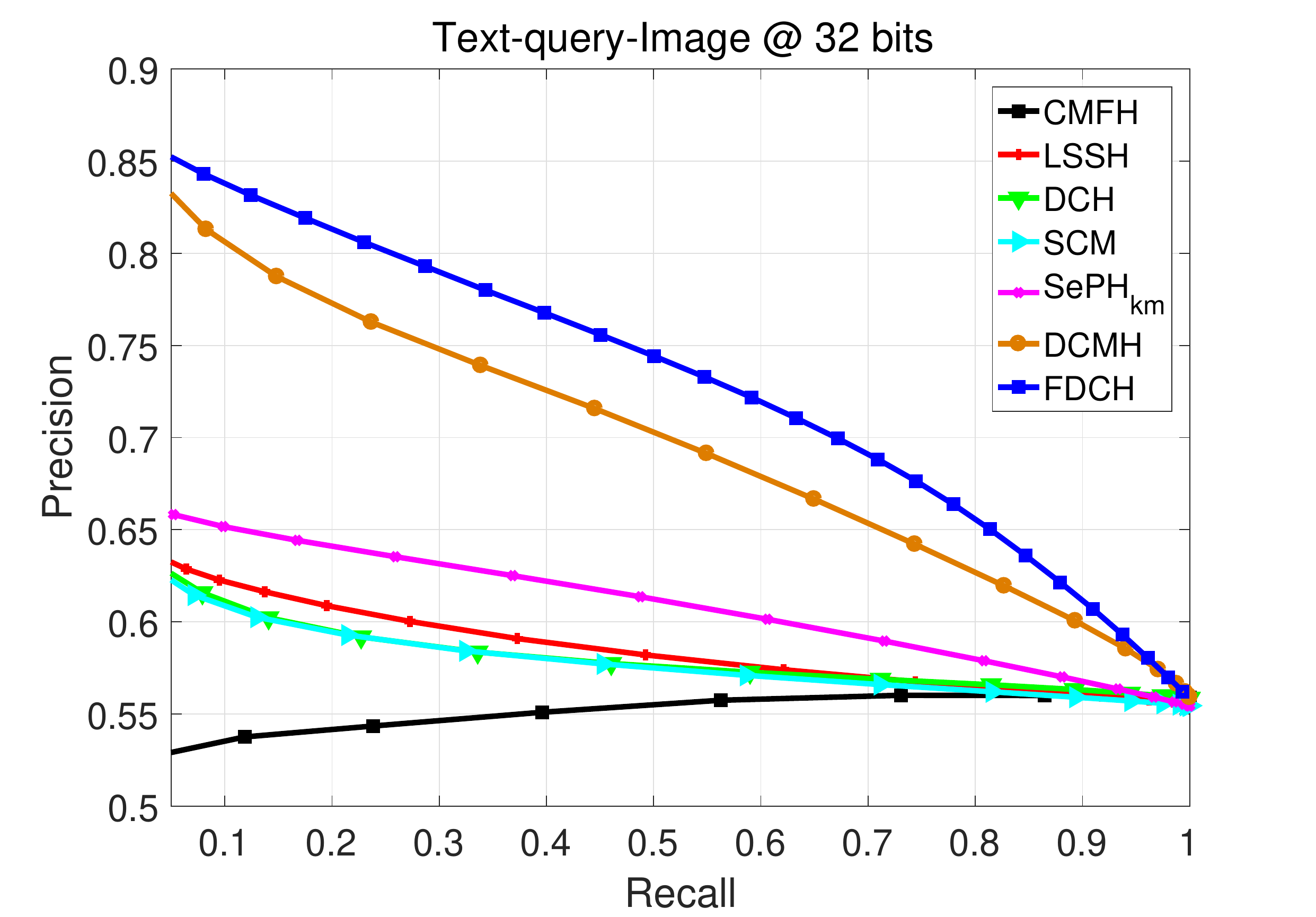}}%\hspace{-5mm}
}
\mbox{
\subfigure{\includegraphics[width=70mm]{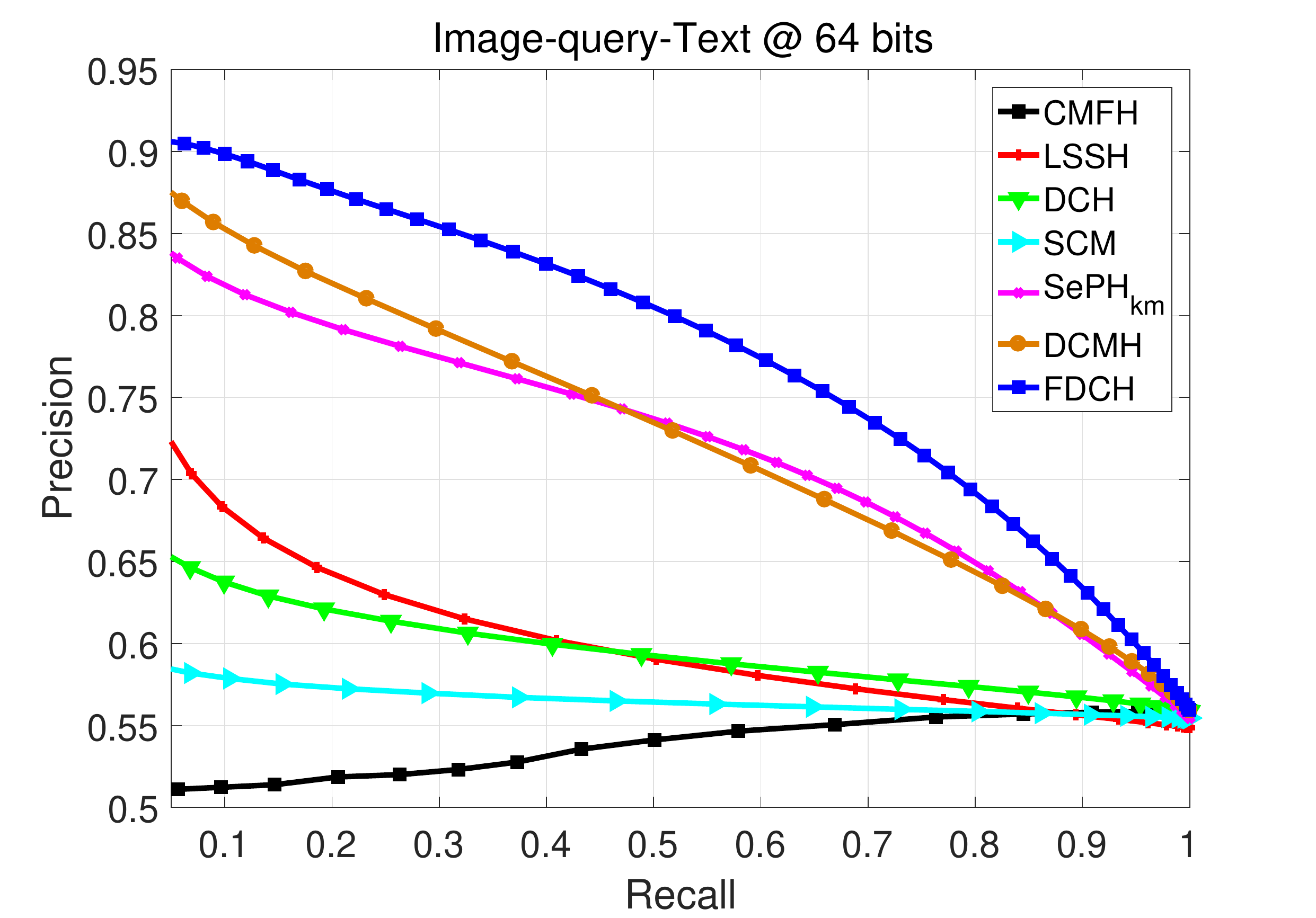}}%\hspace{-5mm}
}\hspace{-4mm}\mbox{
\subfigure{\includegraphics[width=70mm]{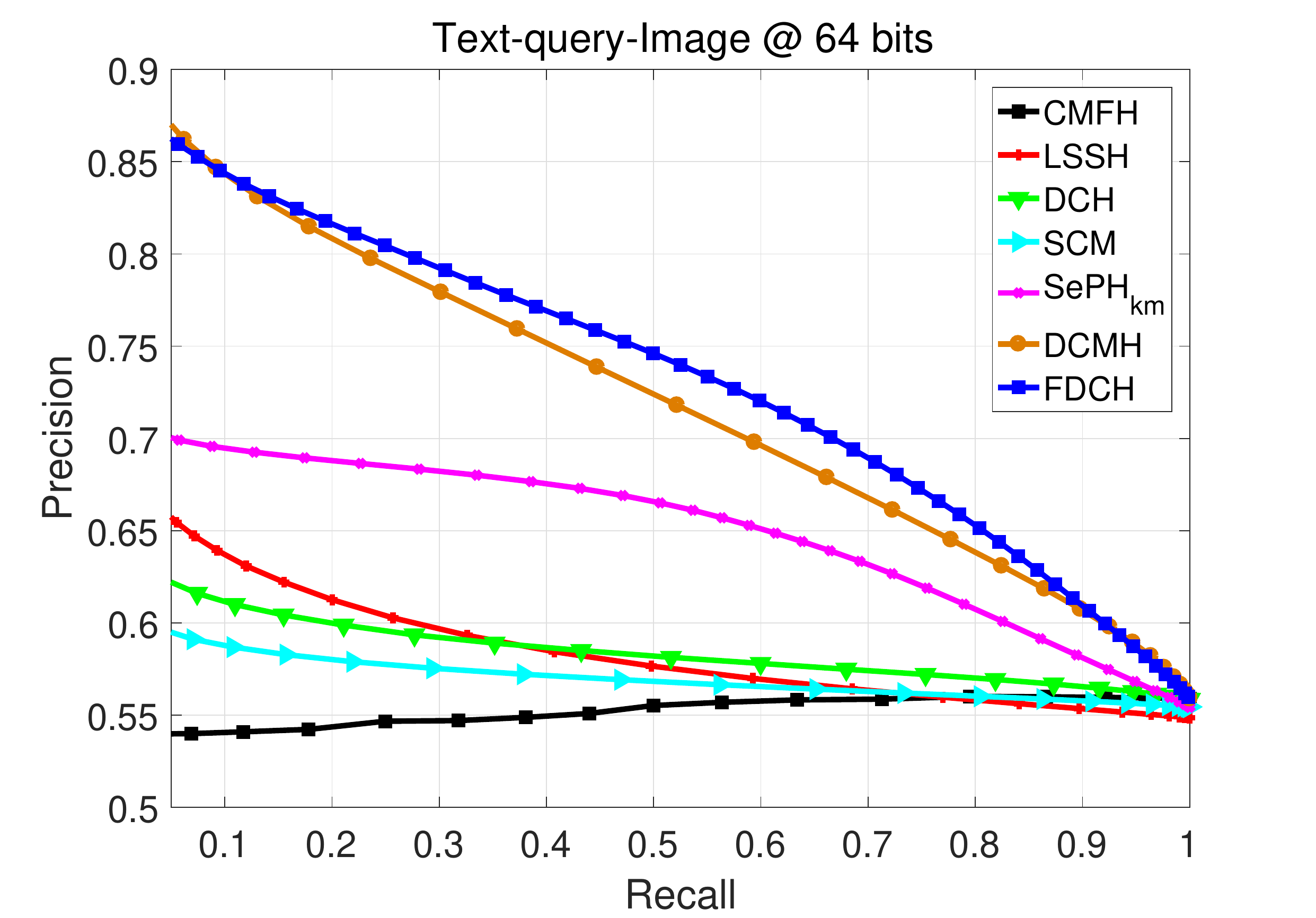}}%\hspace{-5mm}
}
%\vspace{-6mm}
\caption{\emph{Precision-Recall} curves on MIRFLICKR-25K.}
\label{fig:Figure2}
\vspace{-3mm}
\end{figure*}

\subsection{Experimental Settings}
\label{sec:4.2}
\textbf{Baselines}: In experiments, we compare our FDCH with six state-of-the-art cross-modal hashing methods.
%Five shallow-structure-based methods: LSSH \cite{LSSH}, CMFH \cite{CMFH}, DCH \cite{DCH}, SCM \cite{SCM}and SePH$_{km}$ \cite{SePH}.
%Two deep-structure-based methods: DCMH \cite{DCMH} and PRDH \cite{PRDH}.
They can be divided into two categories: LSSH \cite{LSSH}, CMFH \cite{CMFH} and DCH \cite{DCH} are supervised methods; SCM \cite{SCM}, SePH$_{km}$ \cite{SePH} and DCMH \cite{DCMH} are unsupervised methods. For fair comparison, at the $fc7$ from the initial CNN-F network used by our method, we extracted the CNN feature for the shallowed baselines.
%The parameters in all baselines are carefully set according to the suggestions of the original papers.

%\textbf{Evaluation}: We follow \cite{SSAH,DCMH} to evaluate the performance of all methods through two classic retrieval protocols: Hamming ranking and Hash lookup.
%The Hamming ranking protocol ranks the instances from the retrieval set according to the Hamming distance between these instances and the given query, in ascending order.
%The Hash lookup protocol returns all instances within a certain Hamming radius away from a given query.
%Correspondingly, in the experiment, we use two evaluation matrices: mean Average Precision (mAP) and precision-recall (PR) curve to measure the accuracy of Hamming ranking and hash lookup protocol, respectively.

\textbf{Implemental Details}: Our FDCH is implemented under the MatConvNet framework. For image network, we initialize the first seven layers using the CNN-F network pre-trained on the ImageNet dataset. All parameters for other deep neural networks in the FDCH model are randomly initialized.
For the text network with two fully-connected layers, we set the dimension as $[8192 \rightarrow 2500]$ on the MIRFLICKR-25K, and on the NUS-WIDE we set the dimension as $[8192 \rightarrow 1000]$ for 16 bits, and set to $[8192 \rightarrow 600]$ for 32 and 64 bits.
For the fusion network which combines the outputs of image and text networks in pairs, we set the dimension of its fully-connected layers as $[4096 \rightarrow k]$ on all datasets.\vspace{-4mm}
%In the experiment, we empirically set the values of the parameters $\left\{ {\lambda ,\eta ,\gamma ,\beta ,\alpha } \right\}$ to be 1 due to the limitation of time.
%The learning rate is chosen from ${10^{ - 1.5}}$ to ${10^{ - 3}}$.
%Besides, we fix the mini-batch size to be 128, and set iteration number of the outer-loop in Algorithms 1 and 2 to be 500.\vspace{-3mm}
\subsection{Performance Comparisons}
\label{sec:4.3}
\begin{table}[]
\small
\caption{mAP scores of different variants of FDCH on MIRFLICKR-25K with 16-bit code length.}
\vspace{-1mm}
\centering
%\begin{tabular}{|p{16mm}<{\centering}|p{14mm}<{}|p{9mm}<{\centering}|p{9mm}<{\centering}|p{9mm}<{\centering}|}
\resizebox{0.6\columnwidth}{!}{
\begin{tabular}{|c|l|l|l|l|}
\hline
\multicolumn{4}{|c|}{\emph{\textbf{MIRFLICKR-25K}}} \\ \hline
%\multicolumn{5}{|c|}{\emph{\textbf{NUS-WIDE}}} \\ \hline
 \multirow{1}{*}{Task}         & FDCH      & FDCH-I         & FDCH-II   \\
 \hline\multirow{1}{*}{\begin{tabular}[c]{@{}c@{}}I $\rightarrow$ T\end{tabular}}
         & \textbf{0.7762}       & 0.7666         & 0.7379          \\
 \hline\multirow{1}{*}{\begin{tabular}[c]{@{}c@{}} T $\rightarrow$ I\end{tabular}}
         & \textbf{0.7504}       & 0.7310          & 0.6834          \\ \hline
\end{tabular}
}
\label{table1}
\vspace{-6mm}
\end{table}
%\subsubsection{Hamming Ranking.} %, while the DCH, SCM and DCMH generate hash codes for the gallery set through the learned hash functions.%and use it as the gallery set
%On the MIRFLICKR-25K, compared with the shallow methods CMFH, LSSH, SCM and DCH, our method achieved a 10% increase in MAP scores on both retrieval tasks .
%\label{sec:4.31}
\textbf{Hamming Ranking} \cite{DCMH}: Table \ref{table2} shows the mAP scores of our FDCH and the compared methods with CNN-F features on MIRFLICKR-25K and NUS-WIDE, respectively.
In the cross-modal retrieval task, LSSH, CMFH, SePH$_{km}$ and our FDCH both learn the unified hash codes. As can be seen from Table \ref{table2}, our method achieves the best performance for two different retrieval tasks on MIRFLICKR-25K. This is because our approach can effectively capture the correlation of multi-modal data through the nonlinear transformation from the fusion network. On the NUS-WIDE with complex semantic content, our FDCH also obtains superior performance compared to other methods, as shown in Table \ref{table2}. The results demonstrate that FDCH can fully exploit the label information to learn more discriminative hash code.

%Since the source codes for the PRDH \cite{PRDH} are not publicly available, we compare directly with the results of PRDH by using the same experimental setup as its original paper. The mAP results on MIRFLICKR-25K and NUS-WIDE are shown in Table \ref{table2}. PRDH exploits both inter-modal and intra-modal pair-wise embedding constraints to enforce the hash codes. Like our approach, it also uses the CNN-F model for image networks. However, as can be seen from Table \ref{table2}, the map scores of our FDCH are 10\% higher than PRDH on the I2T task over three code lengths, and we also achieves better performance on the T2I task. This is because our FDCH can capture the nonlinear correlations of different modalities more effectively through the fusion network.

%To further demonstrate the effectiveness of our FDCH on large-scale datasets. On MIRFLICKR-25K dataset, we utilize all the samples in the database as training data to learn the hash functions. The results in Table \ref{table5} show that our approach on large datasets still achieved promising performance.

\begin{figure*}
\centering
\mbox{
\subfigure{\includegraphics[width=70mm]{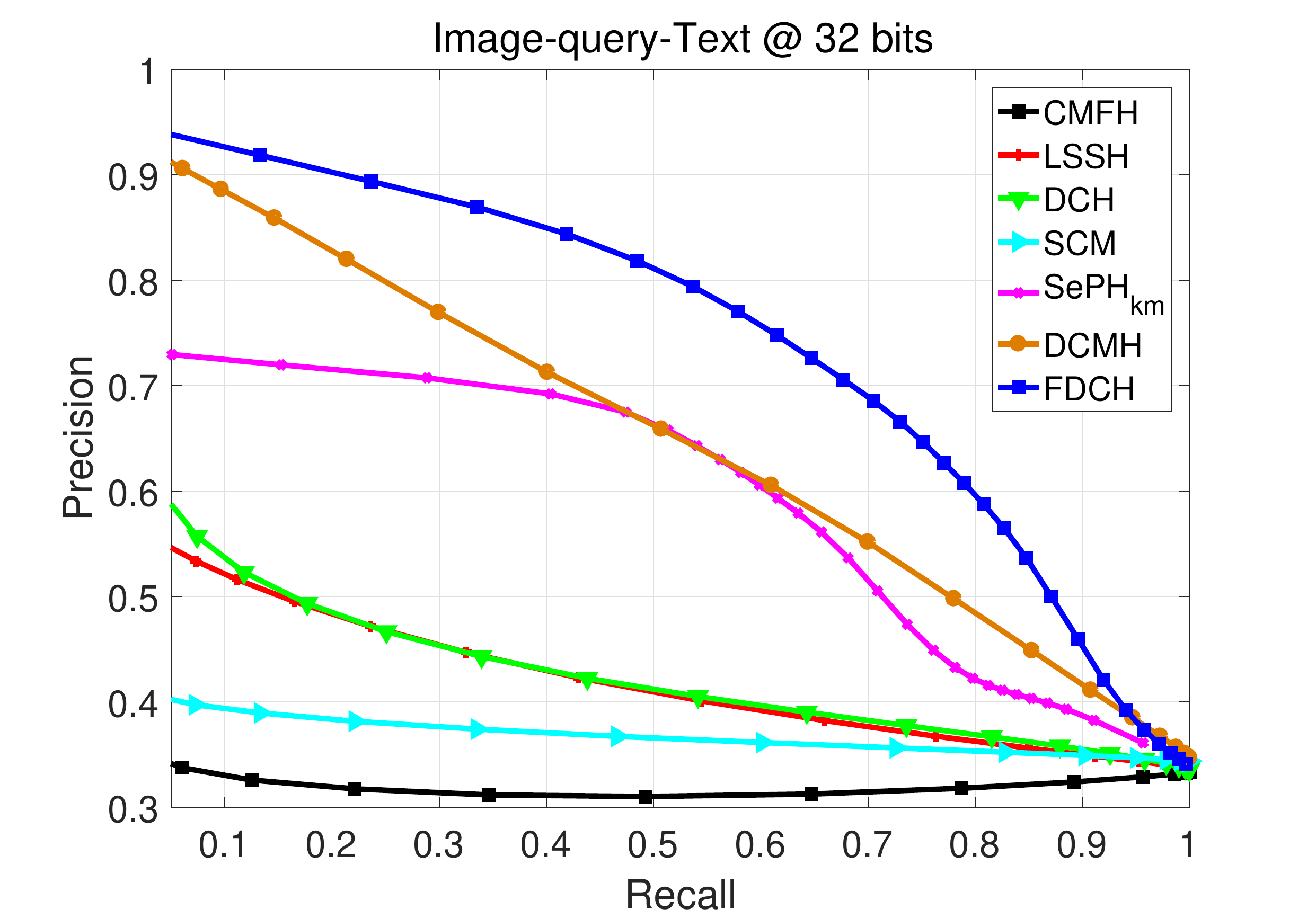}}% \hspace{-5mm}
}\hspace{-4mm}\mbox{
\subfigure{\includegraphics[width=70mm]{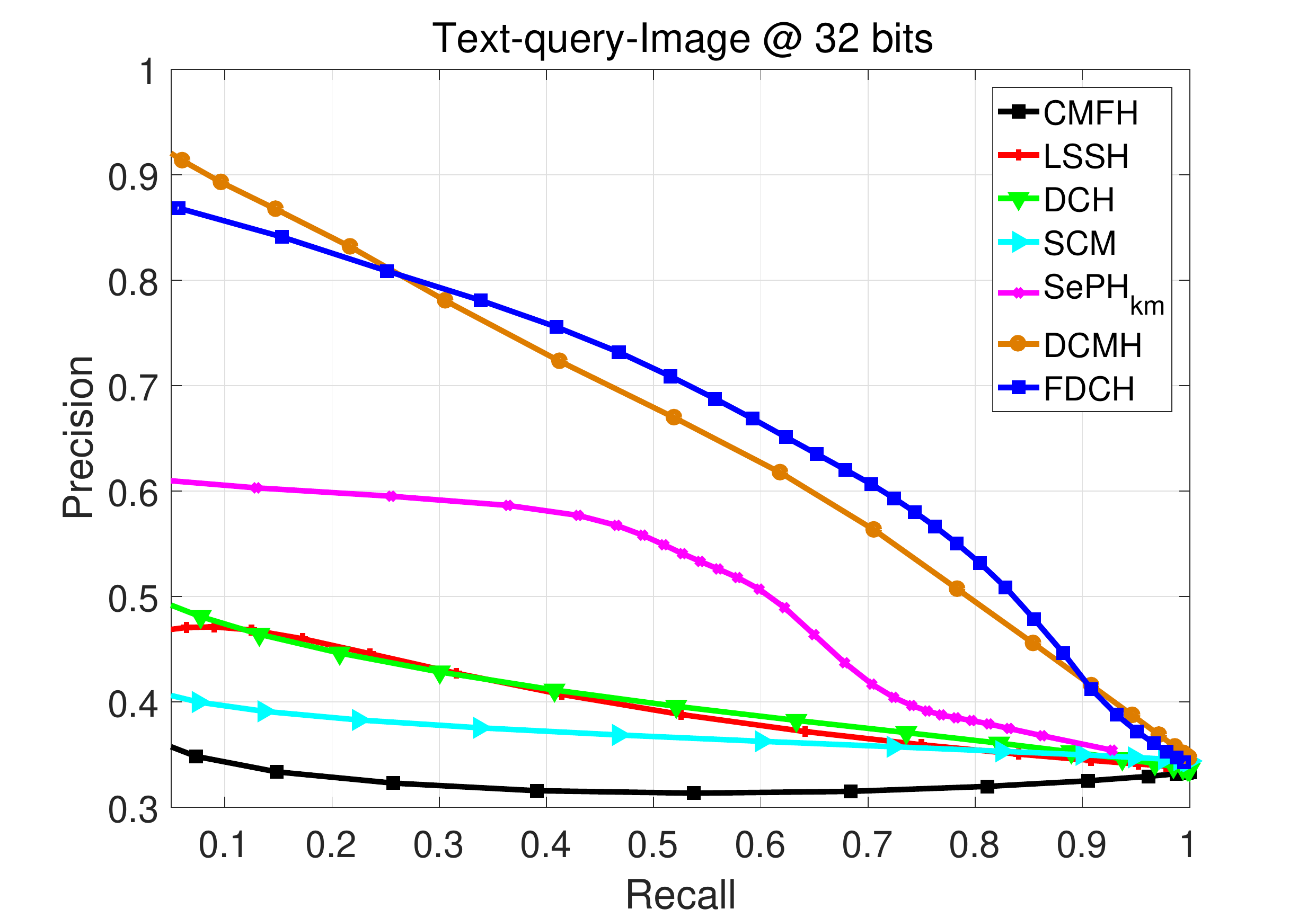}}%\hspace{-5mm}
}\hspace{-4mm}\mbox{
\subfigure{\includegraphics[width=70mm]{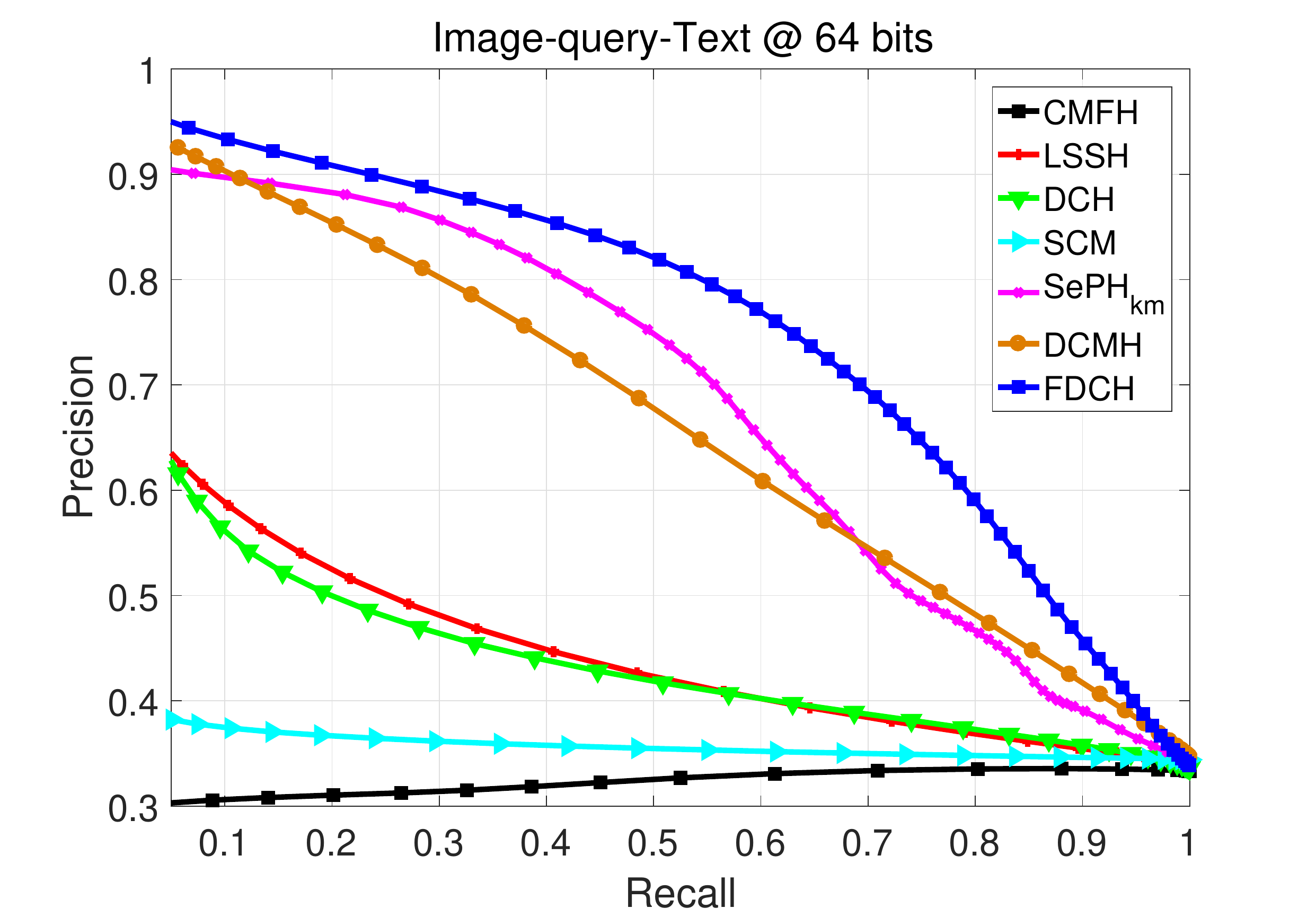}}%\hspace{-5mm}
}\hspace{-4mm}\mbox{
\subfigure{\includegraphics[width=70mm]{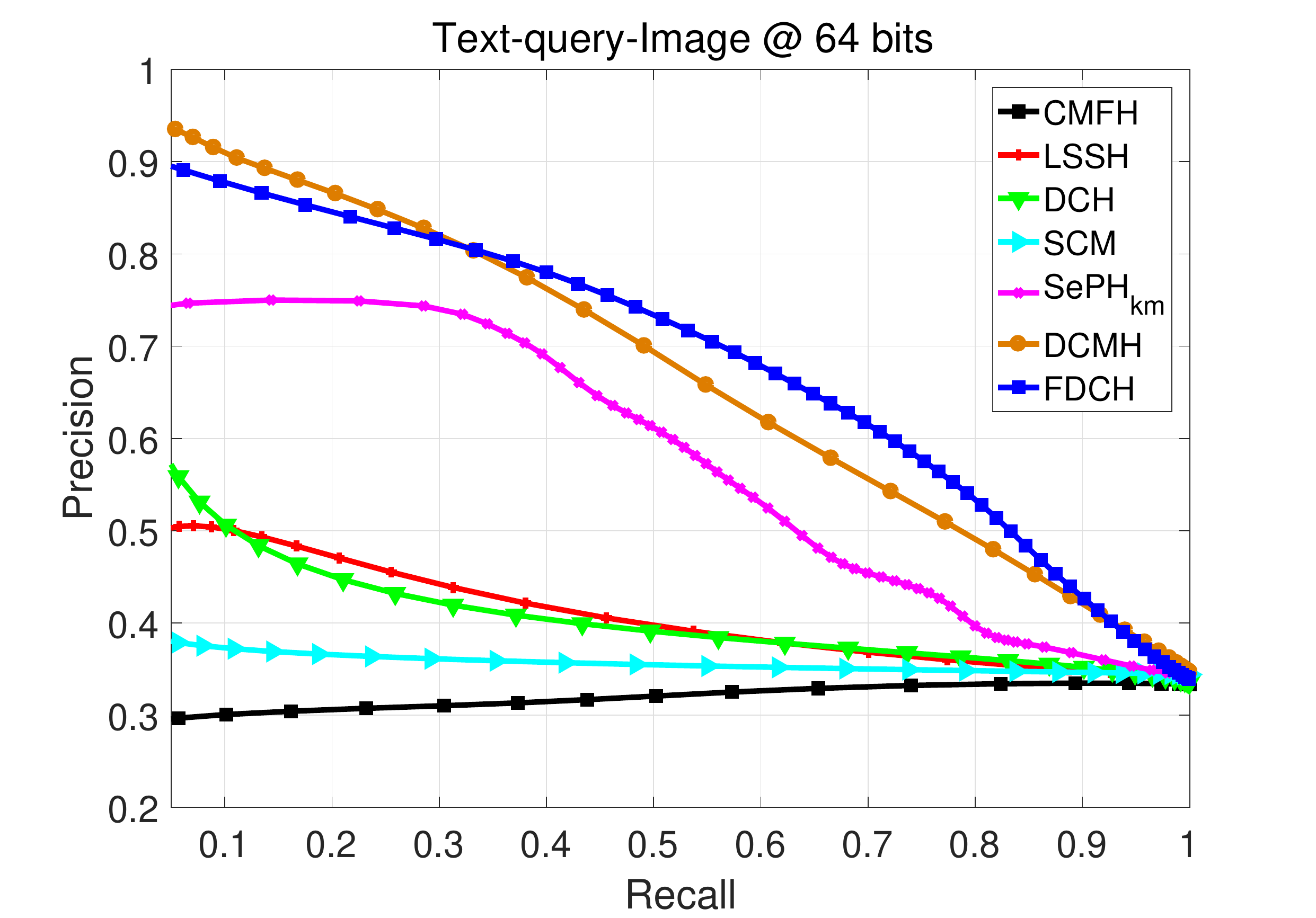}}%\hspace{-5mm}
}
%\vspace{-6mm}
\caption{\emph{Precision-Recall} curves on NUS-WIDE.}
\label{fig:Figure3}
%\vspace{-6mm}
\end{figure*}

%\subsubsection{Hash Lookup.}
%\label{sec:4.32}
\textbf{Hash Lookup} \cite{DCMH}: Figure \ref{fig:Figure2} and \ref{fig:Figure3} show the precision-recall curves by Hash Lookup. We plot the precision-recall curve by varying the Hamming radius from 0 to $k$ with a step size of 1. Figure \ref{fig:Figure2} presents the precision-recall curves with 32 and 64 code lengths for different retrieval tasks on MIRFLICKR-25K. The precision-recall curves on NUS-WIDE are shown in Figure \ref{fig:Figure3}. It can be seen that the performance of our FDCH is significantly better than baselines.

Further, we design two variants of our method to evaluate the performance. FDCH-I ignores the classification information embedded in the modality-specific network. FDCH-II removes the pair-wise similarity constraint when training modality-specific hash network. Table \ref{table1} shows that the mAP scores of these variants are degraded. The results verify that it is effective to simultaneously embed pair-wise similarity and classification information in modality-specific network under one stream framework.\vspace{-4mm}%, especially for the task of retrieving images with text

\section{Conclusion}
\label{sec:5}
In this paper, we propose a novel Fusion-supervised Deep Cross-modal Hashing (FDCH) approach. Through a fusion network with paired samples as input, FDCH effectively explores the complex correlation between multi-modal data. The learned unified hash codes enhance the similarity between instances with the same semantic. Besides, both pair-wise similarity information and classification information are embedded in modality-specific hash networks under one stream framework, which simultaneously preserves cross-modal similarity and semantic consistency. Experiments demonstrate the superiority of FDCH compared with several state-of-the-art approaches.

\textbf{Acknowledgements.} The work is partially supported by the National Natural Science Foundation of China (Nos. 61772322, 61572298, 61802236, U1836216) and the Key Research and Development Foundation of Shandong Province (Nos. 2017GGX10117, 2017CXGC0703)

% References should be produced using the bibtex program from suitable
% BiBTeX files (here: strings, refs, manuals). The IEEEbib.bst bibliography
% style file from IEEE produces unsorted bibliography list.
% -------------------------------------------------------------------------

\footnotesize
\bibliographystyle{IEEEbib}
\bibliography{icme2019template}

\end{document}